\documentclass[aps,pra,twocolumn,superscriptaddress]{revtex4}
\usepackage{graphicx}
\usepackage{amsmath,amssymb}
\begin{document}
\title{Quantum metrology at level anti-crossing}
\author{Luca Ghirardi}
\affiliation{Dipartimento FIM, Universit\`a 
di Modena e Reggio Emilia, I-41125 Modena, Italy}
\author{Ilaria Siloi}
\affiliation{Department of Physics, University of North Texas, 76201 Denton, TX, USA}
\author{Paolo Bordone}
\affiliation{Dipartimento FIM, Universit\`a 
di Modena e Reggio Emilia, I-41125 Modena, Italy}
\affiliation{Centro S3, CNR-Istituto di Nanoscienze, 
I-41125 Modena, Italy}
\author{Filippo Troiani}
\affiliation{Centro S3, CNR-Istituto di Nanoscienze, 
I-41125 Modena, Italy}
\author{Matteo G. A. Paris}\email{matteo.paris@fisica.unimi.it}
\affiliation{Quantum Technology Lab,
Dipartimento di Fisica dell'Universit\`{a} degli Studi di
Milano, I-20133 Milano, Italy}
\affiliation{INFN, Sezione di Milano, I-20133 Milano, Italy}
\date{\today}                                           
% Activate to display a given date or no date
\begin{abstract} 
We address parameter estimation in two-level systems 
exhibiting level anti-crossing and prove that universally optimal strategies for
parameter estimation may be designed, that is, we may find a parameter independent 
measurement scheme leading to the ultimate quantum precision independently on the 
nature and the value of the parameter of interest. Optimal estimation may be achievable also at high temperature depending on the structure of the two-level Hamiltonian. Finally, we show that no improvement is achievable by dynamical strategies and discuss examples of applications.
\end{abstract}
\maketitle
\noindent
\section{Introduction}
The avoided level-crossing theorem \cite{vnw,raz72}, often referred to as the 
level anti-crossing theorem, describes a characteristic phenomenon occurring 
in systems with a parameter dependent Hamiltonian. It states that if 
the Hamiltonian depends on $n$ real parameters, then the eigenvalues 
cannot be degenerate, apart from a $(n-2)$-dimensional manifold in the 
parameter space. For a system Hamiltonian depending on a single parameter 
this means that the eigenvalues cannot {\em cross} at all as a function
of the parameter itself. Level anti-crossing, also referred to as 
level repulsion, plays a relevant role in 
several branches of quantum physics and chemistry 
\cite{lac1,wok81,lac2,hei90,prc1} and 
frequently arises in the study of condensed matter systems.
In systems with parameter dependent Hamiltonian, and thus anticrossing, 
small perturbations to the parameter may induce relevant changes 
in the system ground state \cite{zphy,zph1}, which are possibly reflected in large
variations of some accessible observable. Level anti-crossings, which is also
connected to creation of resonances \cite{res2,res1} and the onset of chaos 
\cite{cha0,cha3,cha2,cha1}, may thus represent a resource for the characterization 
of Hamiltonians and/or the estimation of parameters \cite{molma,wic16}.
\par
In this paper, we address in details metrological applications of level 
anti-crossing and show that {\em universally optimal} strategies for 
parameter estimation may be designed. By this terminology we mean that 
we may find a parameter 
independent measurement scheme leading to the ultimate quantum precision 
independently on the nature and the value of the parameter of interest. 
In particular, we address quantum estimation for parameter dependent 
two-level Hamiltonians \cite{pars0,pars1,pars5,pars4,pars6,pars2,pars3} 
and show analytically that universal optimal estimation 
is achievable, that is, the ultimate precision permitted by quantum mechanics 
may be obtained by a class of parameter-independent measurement schemes. This 
is of metrological interest since it is often the case that the description and 
the dynamics of a metrological system may be restricted to the effective 
two-level system made of  its two lowest energy levels. 
\par
The paper is structured as follows. In Section \ref{s:sys} we introduce 
notation and the basic tools to analyze two-level systems with parameter
dependent Hamiltonian. In Section \ref{s:lqe} we discuss the ultimate 
quantum bounds to precision of parameter estimation, whereas in Section 
\ref{s:pvm} we show how those limits may be achieved by parameter independent 
measurement schemes, including estimation at finite temperature. We also show
that dynamical estimation strategies cannot improve performances. 
Section \ref{s:exa} is devoted to some examples and Section \ref{s:out}
closes the paper with some concluding remarks.
\section{The system}\label{s:sys}
Let us consider a two-level system governed by a parameter 
dependent Hamiltonian of the form
\begin{align}\label{hamlam}
{\cal H} & = \left(\begin{matrix} \omega_1 (\lambda) & \gamma(\lambda) \\ 
\gamma^* (\lambda) & \omega_2 (\lambda) \end{matrix}\right)  \,.
\end{align}
The parameter $\lambda$ is the quantity of interest. It is initially 
unknown and we want to estimate its value by performing measurements
on the system. We assume that $\lambda \in \Lambda$ where $\Lambda$ is
a generic subset of the real field. The eigenvalues of ${\cal H}$ 
are given by
\begin{align}
h_\pm (\lambda) & =\omega_0 (\lambda) \pm 
\sqrt{|\gamma(\lambda)|^2  + \Delta^2 (\lambda)}\label{eigh}\\ 
h_+-h- & = 2 \Delta(\lambda) \sqrt{1+|x|^2} \label{dh} \\ 
x & =\frac{\gamma(\lambda)}{\Delta(\lambda)} \notag\,,
\end{align}
where 
\begin{align}
\omega_0 (\lambda) & = \frac12 \left[\omega_2(\lambda) + \omega_1(\lambda)\right]\,,\\
\Delta (\lambda) & = \frac12 \left[\omega_2(\lambda) -\omega_1(\lambda)\right]\,.
\end{align}
Upon looking at Eqs. (\ref{eigh}) and (\ref{dh}) we see that 
$h_-<h_+$, $\forall \lambda$ i.e. no level crossing occurs for any
value of the parameter of interest $\lambda$.
In the following, without loss of generality, we will assume
that $\omega_2(\gamma)>\omega_1(\gamma)>0$, i.e. $\Delta(\lambda) >0$ 
and  $\gamma (\lambda)\in\mathbb{R}$, $\forall \lambda$. In order to 
simplify notation, we also drop the explicit 
dependence on $\lambda$
of $\gamma$ and $\Delta$.
The Hamiltonian may be rewritten as 
\begin{align}
H & = \omega_0 \sigma_0 - \Delta  \sigma_3 + \gamma  \sigma_1 \label{hamm}
\end{align}
where $\sigma_k$, $k=0,..,3$ denote the Pauli matrices and the term 
$\gamma  \sigma_1$ is usually referred to as the {\em transverse} part
of the Hamiltonian.
The projectors over the eigenvectors $|\psi_\pm\rangle$ of ${\cal H}$ may 
be expressed as follows
\begin{align}
P_\pm (\lambda) \equiv |\psi_\pm\rangle\langle\psi_\pm| & = \frac12 \left[\mathbb{I} 
\pm \frac{\Delta\sigma_3 -  \gamma  \sigma_1}{\sqrt{ \gamma^2  + \Delta^2 }} \right] \label{pmp} 
\\ & = \frac12 \left[\mathbb{I} \pm \frac{\sigma_3 - x \sigma_1}{\sqrt{1 + x^2 }} \right]\,,
\end{align}
and are independent on $\omega_0$.
\section{Ultimate bounds to precision}
\label{s:lqe}
In order to gain information about the value of the parameter $\lambda$, 
which may not correspond to an observable, one performs repeated 
measurements on the system and suitably processes data. The optimal measurement 
corresponds to the spectral measure of the so-called symmetric logarithmic 
derivative (SLD) $L_\lambda$, which is defined by the Lyapunov-like equation
\begin{align}\label{sldef}
\partial_\lambda\rho_\lambda = \frac12 \left(L_\lambda\rho_\lambda + 
\rho_\lambda L_\lambda \right)\,,
\end{align}
being $\rho_\lambda$ the (parameter-dependent) state of the system \cite{hlb,lqe}.
At zero temperature the system stays in its ground state and the SLD 
reduces to 
\begin{align}
L_\lambda & = 2 \left(|\partial_\lambda\psi_-\rangle\langle\psi_-|
+|\psi_-\rangle\langle\partial_\lambda\psi_-| \right) \\ & = 
\frac{\partial_\lambda x}{(1+x^2)^\frac32} (\sigma_1 - x \sigma_3)\label{sld}\,.
\end{align}
By measuring $L_\lambda$ on repeated preparations of the system one 
collects data and then builds an estimator for the unknown quantity $\lambda$, i.e.
a function $\hat\lambda(\chi)$ of the data sample $\chi=\{x_1,x_2,...,x_{\hbox{\tiny M}}\}$
that returns the value of the parameter when averaged over data.
The precision of the overall estimation strategy corresponds to 
the variance of the estimator. An efficient estimator (e.g. the 
maximum-likelihood estimator or the Bayesian one) has variance 
saturating the quantum Cramer-Rao bound 
$\hbox{Var}\hat\lambda = 1/M H(\lambda)$ where 
$M$ is the number of measurements and
$H(\lambda)=\hbox{Tr}\left[\rho_\lambda\,L_\lambda^2\right] = \langle 
\psi_- | L_\lambda^2 |\psi_-\rangle$ is the 
so-called Quantum Fisher information (QFI)
\cite{e1,e2,e3,e4,e5,e6,e7,e8,e9,e10,e11,e12,e13,e14,e15,e16}.
Notice that the optimal measurement, and the corresponding precision,
do explicitly depend of the value of $\lambda$.
Using Eq. (\ref{sld}) one has
\begin{align}
H(\lambda) & = \frac{\left(\partial_\lambda x\right)^2}{(1+x^2)^2}\,,\\
& = \frac{\big(\Delta  \partial_\lambda\gamma -
\gamma \partial_\lambda\Delta\big)^2 }{\gamma^2
+ \Delta^2}\label{qfi0} \\
& = 16\, \left(\frac{\Delta}{h_+-h_-}\right)^4 \left(\partial_\lambda x \right)^2\,,
\end{align}
where the last expression well illustrates the connections with level 
anti-crossing. The same result may be obtained from the expression of 
the QFI in term of the ground state fidelity \cite{fid1,fid2,fid3,wim11,geoNG}, 
i.e. 
\begin{align}
H(\lambda) & = \lim_{\delta\lambda\rightarrow 0} 4\, \frac{1-\big|\langle \psi_- 
(\lambda+\delta\lambda)|\psi_-(\lambda)\rangle\big|}{\delta\lambda^2}
\,.\label{qfi1}
\end{align}
Using Eq. (\ref{qfi1}) it may be proved that the QFI for 
{\em any} set of superposition states of the form 
$|\psi_\theta\rangle=\cos\theta |\psi_-\rangle 
+ \sin\theta |\psi_+\rangle$ is equal to the ground state one.
\par
As it is apparent from the above expressions, the QFI, and in 
turn the precision of any estimation scheme, does not depend on $\omega_0$. 
 Notice also that if either 
$\Delta(\lambda)=0$ or $\gamma(\lambda)=0$, $\forall \lambda$ then $H(\lambda)=0$ 
and no  estimation strategy is possible. This behaviour may be understood
by looking at Eq. (\ref{pmp}), which shows that for $\Delta(\lambda)=0$ 
or $\gamma(\lambda)=0$ the eigenstates of the system become 
$P_\pm(\lambda) = \frac12 [\mathbb{I} \mp \hbox{sgn} (\gamma)\, \sigma_1]$ or
$P_\pm(\lambda) = \frac12 [\mathbb{I} \pm \hbox{sgn} (\Delta)\, \sigma_3]$ respectively. 
In both cases the ground state is independent on $\lambda$ (except for
the crossing values), and no information may be
gained by performing measurements on the system.
\section{Universally optimal estimation by projective measurements}
\label{s:pvm}
Since the
SLD does depend on the unknown value of the parameter a question
arises on whether the ultimate precision may be actually 
achieved without a priori information. As we will see, universal
estimation based on a single detector implementing a parameter
independent measure may be indeed obtained. 
\par
\begin{figure}[h!]
\includegraphics[width=0.95\columnwidth]{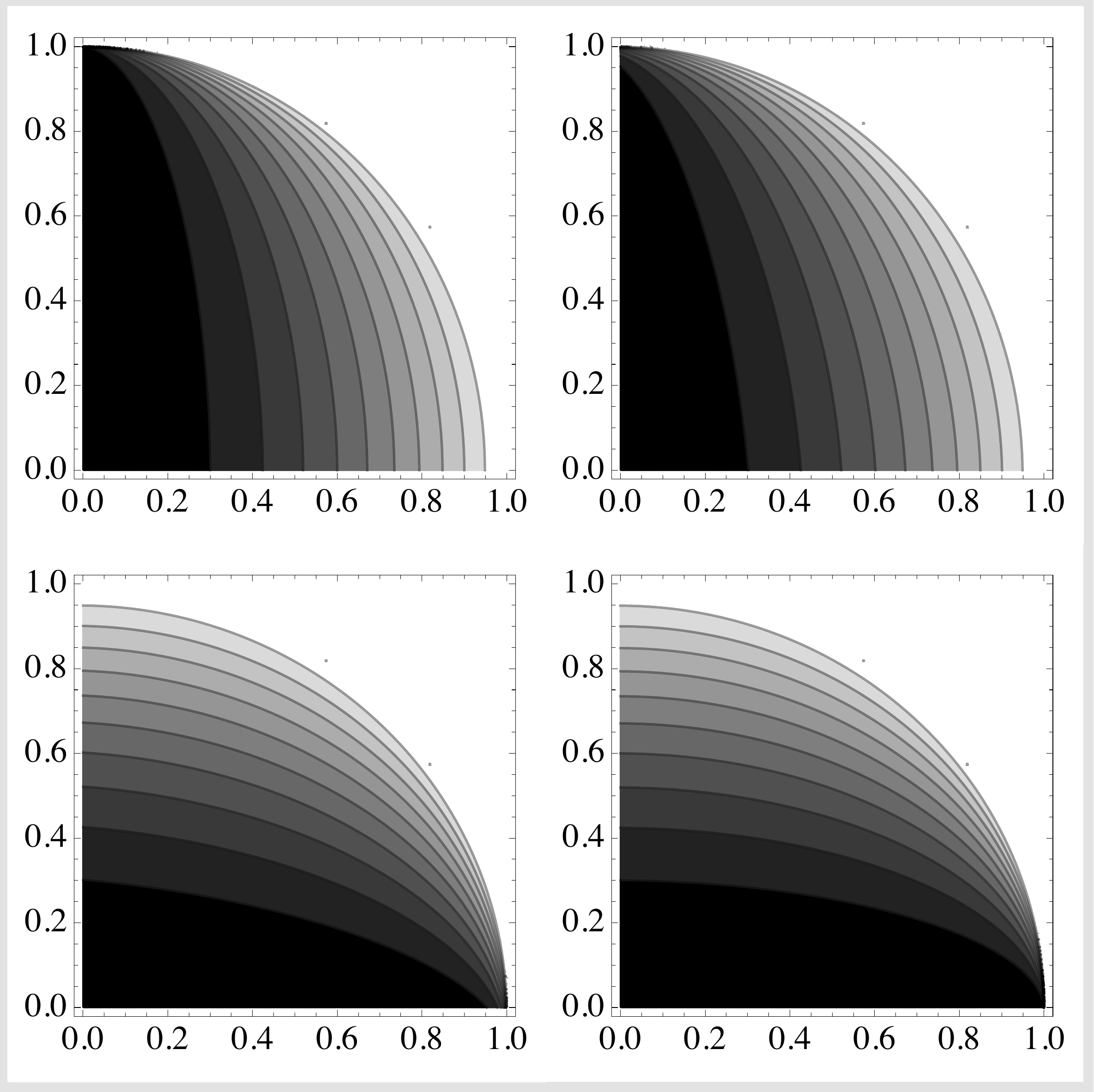}
\caption{Density plot of $g_\lambda (r_1,r_3)$ at fixed values of $x$ 
as a function of $r_1$ and $r_3$. From top left to bottom right we show 
$g_\lambda (r_1,r_3)$ for $x=0.01, 0.1, 10, 100$ respectively.
The function is defined only in the region $r_1^2+r_3^2\leq1$. Darker regions
correspond to lower values of $g$.}
\label{f:gg}
\end{figure}

A generic (projective) measurement on a two-level system is described
by the operatorial measure $\{\Pi, \mathbb{I}-\Pi\}$ where 
\begin{align}
\Pi=\frac12 (\mathbb{I} + \mathbf{r}\cdot{\bf\sigma})\,,
\label{pidef}
\end{align} 
and $|\mathbf{r}|=1$.
The distribution of the two possible outcomes is governed by the
probability
\begin{align}
q_0(\lambda)& \equiv q(\lambda) = \hbox{Tr}[\rho_\lambda\,\Pi] = \frac12 \left( 1+ 
\frac{x r_1 - r_3}{\sqrt{1+x^2}} \right) \label{q00}\\ 
q_1(\lambda)&= \hbox{Tr}[\rho_\lambda\,(\mathbb{I} - \Pi)]=1-q(\lambda)
\,.
\end{align}
The variance of any estimator is now bounded by the classical Cram\'er-Rao bound 
$\hbox{Var}\hat\lambda \geq 1/M F(\lambda)$ \cite{crb} and efficient estimators are 
those saturating the bound, where $F(\lambda)$ is the Fisher information 
of the distribution 
$q_k(\lambda)$, i.e. 
\begin{align}
F(\lambda) & = \sum_k \frac{\left(\partial_\lambda q_k\right)^2}{q_k}=
\frac{\left(\partial_\lambda q\right)^2}{q(1-q)} \\ 
&=  \, H(\lambda)\, g_\lambda (r_1,r_3) \label{fhx}\,.
\end{align}
where 
\begin{align}
g_\lambda(r_1,r_3) & \equiv g (x,r_1,r_3)  \notag \\ & = 
\frac{(r_1 + x r_3)^2}{1+x^2-(x r_1-r_3)^2} \label{fhg}\,.
\end{align}
As expected from the quantum Cram\'er-Rao theorem we have $F(\lambda) \leq H(\lambda)$, 
i.e. $g_\lambda (r_1,r_3) <1$ $\forall \lambda, r_1, r_3$ (see Fig. \ref{f:gg}).
On the other hand, we have equality, $F(\lambda) = H(\lambda)$, either for $r_1=1$
and $r_3=0$ or for $r_1=0$ and $r_3=1$, i.e. by measuring either $\sigma_1$ or
$\sigma_3$ on the two-level system. In addition, if $r_2=0$ we have 
$r_1=\sqrt{1-r_3^2}$ and $F(\lambda) = H(\lambda)$, $\forall r_3$, i.e. any observable
of the form $\sigma_\theta=\sigma_1 \sin\theta + \sigma_3\cos\theta$ leads
to optimal estimation. 
\par
In other words, Eq. (\ref{fhx}) and the following 
arguments show that universal optimal estimation, 
i.e. maximum precision for any value of $\lambda$, may be achieved by parameter 
independent measurements.
\par
Let us now discuss robustness of the estimation strategy. The discussion 
above has shown that the optimal (projective) measurement corresponds to 
the choice $r_2=0$ and any pair $(r_1,r_3)$ satisfying $r_1^2+r_3^2=1$ for 
the expression of the operator measure $\Pi$ in Eq. (\ref{pidef}). 
On the other hand, some class of pairs may be better than others in 
practical implementation, depending on the relative values of 
$\gamma(\lambda)$ and $\Delta(\lambda)$, i.e. the value of $x$. 
Indeed, as it is apparent from the upper panels of Fig. \ref{f:gg}, 
if $\gamma (\lambda) \ll \Delta (\lambda)$ in the whole range of 
variation of $\lambda$, then $x\ll 1$ and $F(\lambda) \simeq H(\lambda)$ 
also if some imperfections lead to the measurement of a slightly perturbed 
observable corresponding to $r_2 \gtrsim 0, r_3 \gtrsim 0, r_1 \lesssim 1$, 
rather than the optimal ideal one $\sigma_\theta$. The situation is reversed
if $\gamma(\lambda) \gg \Delta (\lambda)$ in the whole range of variation 
of $\lambda$, see the lower panels of Fig. \ref{f:gg}, and 
also from the symmetry $g(x,r_1,r_3)=g(1/x,r_3,r_1)$ of the function $g$.
\par
\subsection{Estimation at finite temperature}
If the system is not at zero temperature
the equilibrium state is given by 
\begin{align}
\rho_{\lambda\beta} =  p_+ P_+ + p_- P_- \,,
\end{align}
where $\beta$ is the inverse temperature and 
the projectors $P_\pm$ over the eigenvectors $|\psi_\pm\rangle$ are given 
in Eq. (\ref{pmp}). The probabilities $p_\pm = e^{-\beta h_\pm}/Z $ are obtained
from the eigenvalues $h_\pm$ of Eq. (\ref{eigh}) and from the partition function 
$Z = e^{-\beta h_+ } + e^{-\beta h_- }$, i.e.
\begin{align}
Z = 2\, e^{-\beta \omega_0 } \cosh \left(\beta \sqrt{\gamma^2  + \Delta^2 }\right) \,.
\end{align}
Using the above expressions we arrive at
\begin{align}
p_\pm = \frac12  \left[1 \pm \tanh \left(\beta \sqrt{\gamma^2  + 
\Delta^2 }\right) \right]\,,
\end{align}
and, in turn, to
\begin{align}
\rho_{\lambda\beta} = \frac12 \left[\mathbb{I} - \tanh \left(\beta \Delta \sqrt{1 + 
x^2}\right)\frac{\sigma_3 - x \sigma_1}{\sqrt{1 + x^2 }} \right]
\,,
\end{align}
which is a mixed state with purity
\begin{align}
\mu_{\lambda\beta}=\hbox{Tr} \left[\rho_{\lambda\beta}^2\right] = \frac12 \left[
1+\tanh^2 \left(\beta \Delta \sqrt{1 + 
x^2}\right)\right]\,.
\end{align}
The quantum Fisher information is now given by sum of two terms $H_\beta(\lambda) = 
H_C(\lambda) + H_Q(\lambda)$ usually referred to as the {\em classical} and 
{\em quantum} part of the QFI. The classical part corresponds to the Fisher 
information of the spectral eigenmeasure  i.e. 
\begin{align}
H_C(\lambda) & = \frac{\left(\partial_\lambda p_+\right)^2}
{p_+p_-}  \notag \\ & = 
\frac{\left(\gamma\partial_\lambda\gamma+
\Delta\partial_\lambda\Delta\right)^2}{\gamma^2  + 
\Delta^2 }\, k_C(\beta,\lambda)  \\
k_C(\beta,\lambda) &= \frac{\beta^2}{\cosh^2
\left(\beta \sqrt{\gamma^2 + 
\Delta^2 }\right)} \,.
\end{align}
The quantum part $H_Q$ take into account the contribution coming from 
the dependence of the eigenvectors on $\lambda$, we have
\begin{align}
H_Q(\lambda) & = 2 \sum_{j,k=\pm} |\langle \psi_j| \partial_\lambda \psi_k\rangle|^2\, \frac{(p_j-p_k)^2}{p_j+p_k} \notag \\ 
& = H_0(\lambda)\, k_Q(\beta,\lambda)\\
k_Q(\beta,\lambda)&=\tanh^2 \left(\beta \sqrt{\gamma^2 + 
\Delta^2 }\right)\,,
 \end{align}
where $H_0(\lambda)$ is the zero-temperature QFI reported in Eq. (\ref{qfi0}). 
In the limit of low temperature we have
\begin{align}\label{lowT}
k_Q(\beta,\lambda)  \stackrel{\beta\gg 1}{\simeq} 1  \qquad
k_C(\beta,\lambda)  \stackrel{\beta\gg 1}{\simeq} 0\,,
\end{align}
whereas for high temperature one may write
\begin{align}\label{highT}
k_Q(\beta,\lambda)  \stackrel{\beta\ll 1}{\simeq} \beta^2 (\gamma^2 + 
\Delta^2 )  \qquad
k_C(\beta,\lambda)  \stackrel{\beta\ll 1}{\simeq} \beta^2 \,.
\end{align}
Eqs. (\ref{lowT}) and (\ref{highT})  say that the quantum part $H_Q$ dominates
in the low temperature regime, whereas for high T the two contributions
are of the same order.
\par
%Let us now introduce a polar representation for the quantities 
%$\Delta$ and $\gamma$, i.e.
%\begin{align}
%\Delta(\lambda)  & = \varrho (\lambda) \cos \theta (\lambda) \notag \\
%\gamma(\lambda) & = \varrho (\lambda) \sin \theta (\lambda)\,.
%\end{align}
%This parametrization amounts to rewrite the Hamiltonian as 
%\begin{align}
%H & = \omega_0 \sigma_0 - \varrho \left (\sigma_3 \cos\theta - \sigma_1 \sin\theta\right) 
%\label{hrt}
%\end{align}
%which may be seen as the Hamiltonian of a system with spin $\frac12$ 
%in an external magnetic field, with $\varrho$ providing information 
%about the intensity of the field and $\theta$ specifying its direction.
%Using this parametrization we may write the QFI as follows
%\begin{align}
%H_\beta (\lambda) = k_C(\beta,\lambda)\, (\partial_\lambda \varrho)^2
%+ k_Q(\beta,\lambda)\, (\partial_\lambda \theta)^2\,,
%\end{align}
%which makes apparent the dependence of the classical/quantum part of the QFI
%on the intensity/direction of the external field.
%\par
Given a generic projective measurement the distribution of
the outcomes is now governed by the quantity
\begin{align}
q_\beta (\lambda) & = \hbox{Tr}[\rho_{\beta\lambda}\,\Pi] = 
\hbox{Tr}[(p_+ P_+ + p_- P_-)\,\Pi] \\ 
&= 1 - p_- + q(\lambda) (2  p_- -1) \\ 
& = \frac12 + \left[q(\lambda)-\frac12\right]\tanh \left[\beta \sqrt{\gamma^2  + 
\Delta^2 }\right] \\
& \stackrel{\beta \gg 1}{\simeq} q(\lambda) 
+ \left[q(\lambda)-\frac12\right] e^{-2 \beta \sqrt{\gamma^2  + 
\Delta^2 }}
\,,
 \end{align}
where $q(\lambda)$ is the zero temperature distribution given in Eq. (\ref{q00}).
The fast convergence of the exponential function ensures that optimal estimation
may be achieved also for finite temperature, provided that $\beta \gtrsim \sqrt{\gamma^2  + 
\Delta^2 }$. In the opposite limit of high temperature, i.e. $\beta \ll 1$ we
may expand the Fisher information up to second order in $\beta$
\begin{align}
F_\beta(\lambda)  = & \frac{\left[\partial_\lambda q_\beta (\lambda)\right]^2}
{q_\beta(\lambda) [1-q_\beta(\lambda)]} \notag \\ 
 = &\left(r_1 \partial_\lambda \gamma -
r_3 \partial_\lambda \Delta \right)^2 \beta^2 + O(\beta^2)
\,.\label{fbe} 
%\\ 
%= &\big[(r_1 \cos\theta + r_3 \sin\theta) \varrho\, \partial_\lambda \theta \notag  \\ 
%&+ (r_1 \sin\theta - r_3 \cos\theta)\, \partial_\lambda \varrho \big]^2 \beta^2 + O(\beta^2)
%\label{fbepolar}
 \end{align}
The Fisher information of Eq. (\ref{fbe}) should be compared to the QFI 
$H_\beta(\lambda) = 
H_C(\lambda) + H_Q(\lambda)$ which, up to second order in $\beta$, reads
as follows   
\begin{align}
H_\beta(\lambda) & = \left[\left(\partial_\lambda \gamma \right)^2 + \left(\partial_\lambda \Delta\right)^2
\right] \beta^2 + O(\beta^2) 
%\\
%& = \left[\varrho ^2 \left(\partial_\lambda \theta \right)^2 + 
%\left(\partial_\lambda \varrho\right)^2
%\right] \beta^2 + O(\beta^2)
\,.
 \end{align}
The two quantities coincides, i.e. universal optimal estimation is achievable 
also at high temperature, when only the transverse, or only the diagonal, 
part of the Hamiltonian does depend on the parameter 
$\lambda$, i.e. if either $\partial_\lambda\Delta=0$ or $\partial_\lambda\gamma=0$. 
In those cases, we have $H_\beta(\lambda)\simeq F_\beta(\lambda)$ up to second order 
by performing a projective measurement with $r_1=1,r_2=r_3=0$, or $r_3=1,r_2=r_1=0$ 
respectively. 
\par
On the other hand, again from the expression in Eq. (\ref{fbe})
one finds that a projective measurement with 
%$r_3/r_1=\tan \theta\equiv \gamma/\Delta$
$r_3/r_1=\gamma/\Delta$ 
or 
%$r_3/r_1=-1/\tan \theta=-\Delta/\gamma$ 
$r_3/r_1=-\Delta/\gamma$
is globally optimal without restrictions 
on the form of the Hamiltonian. In this case, however, the optimal measurement is 
not universal, i.e. it depends on the value of the parameter itself.
\subsection{Dynamical estimation strategies}
One may wonder whether having access to the initial 
preparation of the system may improve precision for some class
of estimation strategies. In fact, general considerations about
unitary families of states suggest the opposite \cite{lqe}, i.e. that no 
improvement may be achieved in this way. In order to prove 
this explicitly for our system, let us now address a  scenario 
in which we are able to initially prepare the system in any 
desired state
\begin{align}
|\psi_{\theta}(0)\rangle = \cos\frac\theta2\, |\psi_-\rangle + e^{i\phi} \sin\frac\theta2\,  |\psi_+\rangle
\,,
\end{align} 
which then evolve according to the Hamiltonian
in Eq. (\ref{hamm}). The evolution operator $U  = 
\exp\left\{ - i H t\right\}$ may be 
written as
\begin{align}
U_t  = &  e^{- i \omega_0 t} \Big[ \cos \left(t \sqrt{\gamma^2+\Delta^2} \right) \sigma_0 \notag  \\ 
& - i t\, \hbox{sinc} \left(t \sqrt{\gamma^2+\Delta^2} \right)\left(\gamma\sigma_1 - 
\Delta\sigma_3\right) \Big]\,,
\end{align}
and the evolved state $|\psi_{\theta}(t)\rangle = U_t |\psi_{\theta}(0)\rangle$ as  
\begin{align}
|\psi_{\theta}(t)\rangle = \cos\frac\theta2\,e^{-i h_- t} |\psi_-\rangle + e^{i\phi} \sin\frac\theta2\, e^{-ih_+ t}  |\psi_+\rangle
\,.
\end{align}
The SLD  
$L_\lambda (t) = 2 \left(
|\partial_\lambda|\psi_{\theta}(t)\rangle\langle\psi_{\theta}(t)|
+
|\psi_{\theta}(t)\langle\partial_\lambda|\psi_{\theta}(t)| \right)$
may be easily evaluated, thanks to the covariant nature of the problem
\begin{align}
L_{\lambda t} =\, & 2\, U_t\, \Big[
|\partial_\lambda|\psi_{\theta}(0)\rangle\langle\psi_{\theta}(0)|
\nonumber\\ &+
|\psi_{\theta}(0)\langle\partial_\lambda|\psi_{\theta}(0)| \Big]\, U_t^\dag
\\ = \, & U_t L_{\lambda 0} U_t^\dag
\,,
\end{align}
where $L_{\lambda 0} \equiv L_\lambda$ is given in Eq. (\ref{sld}).
Finally, we have 
\begin{align}
H_t (\lambda) & =  
\langle\psi_{\theta}(t)|L_{\lambda t}^2|\psi_{\theta}(t)\rangle \notag \\
& =  \langle\psi_{\theta}(0)|L_{\lambda 0}^2|\psi_{\theta}(0)\rangle 
\equiv H_0(\lambda)\,,
\end{align}
where $H_0 (\lambda) \equiv H (\lambda)$ is given in Eq. (\ref{qfi0}).
Notice that the above negative arguments hold when the Hamiltonian is given by 
Eq. (\ref{hamlam}), i.e. it depends on the parameter of interest but it is
time-independent. Improved performances, i.e. more precise 
estimation strategies may be instead 
achieved if the two-level Hamiltonian is explicitly
depending on time \cite{b1,b2,b3,b4}.
\section{Examples}\label{s:exa}
\subsection{Level anti-crossing induced by a perturbation}
Let us consider a two-level system with Hamiltonian 
$H=H_0 + \lambda H_1$ where $H_0$ is the bare Hamiltonian
of the system, $H_1$ represents a perturbation and 
$\lambda$, which is the parameter to be estimated, 
is the perturbation strenght \cite{hei90,mug09}. Without loss of 
generality we assume the following structure
\begin{align}
H_0  = \left(\begin{matrix} \omega & 0 \\ 0 & \omega + \delta \end{matrix}\right) \qquad  
H_1  = R \, \left(\begin{matrix} 0 & 0 \\ 0 & \epsilon \end{matrix}\right) R^{\hbox{\tiny T}} 
\end{align}
where $\delta >0$, $\epsilon >0$ and $R$ is a rotation matrix
$$ 
R  = \left(\begin{matrix} \cos\phi & - \sin\phi \\ \sin\phi & \cos\phi \end{matrix}\right)\,
$$
with $\phi\in [0, \pi/2)$. If $\phi=0$, $R={\mathbb I}$ and the two terms $[H_0,H_1]=0$
commute. In this case, the eigenvalues of $H$ are given by $h_-=\omega$, $h_+=\omega + 
\delta + \lambda \epsilon$ and they are crossing at $\lambda_c= -\delta/\epsilon$. For
$\phi\neq 0$ this degeneracy is removed since the two levels are coupled each other.
The two eigenvalues are now given by
\begin{align}
h_\pm = \omega + \frac12 (\delta + \lambda \epsilon) \pm \sqrt{\delta^2 + 
\lambda ^2 \epsilon^2 + 2 \delta \lambda \epsilon \cos 2\phi}\,,
\end{align}
i.e. we have level anti-crossing, which may be exploited for the precise estimation 
of the perturbation coupling $\lambda$. The QFI may be evaluated using 
Eq. (\ref{qfi0}), where the quantities $\Delta$ and $\gamma$ are now given by
\begin{align}
\Delta & = \delta + \epsilon \lambda \cos 2\phi \\
\gamma & = -\frac12 \epsilon \lambda \sin 2\phi
\,.
\end{align}
The QFI $H(\lambda)$ is maximised for $\phi=\pi/4$, i.e. when $H_0$ and $H_1$
are "maximally non-commuting" and in this case it is given by 
\begin{align}
H(\lambda) = \frac{1}{(1+ y^2 \lambda^2)^2} \quad y = \frac{\epsilon}{2\delta} 
\,,
\end{align}
where, rather intuitively, the dependence on the structure of the Hamiltonian
terms $H_0$ and $H_1$ is summarised by the ratio $\epsilon/\delta$.
\subsection{Driven double-well systems}
It is often the case in condensed matter that double-well systems 
exhibit two lowest-energy levels well separated from 
the next pair by a large gap, i.e. larger than the 
other relevant energies, e.g. the tunnelling energy and the frequency of 
the driving field. In those cases, a two-level approximation describes rather well 
the physics of the system, and the dynamics may be understood in terms of 
the celebrated periodic {\em Rabi Hamiltonian}
$$H=\frac12 \omega_0 \sigma_3 + \lambda \sigma_1 \cos \omega t\,, $$ 
where the coupling $\lambda$ is the quantity to be estimated and $\omega$ 
is the frequency of the driving field, which we assume to be known to the 
experimenter. The model cannot be 
solved exactly \cite{jchem}, since the Hamiltonian is not commuting with itself at 
different times. On the other hand, upon going to the appropriate interaction picture
and neglecting the counter-rotating terms, the system may be described by a two-level
time-independent Hamiltonian \cite{holt93} which, in the relevant subspace. reads as 
follows
\begin{align}
{\cal H}_{\hbox{\tiny eff}}= \left(
\begin{matrix}
\frac12 \Omega & \gamma \\
\gamma & - \frac12 \Omega + 2 \omega
\end{matrix}
\right)\,,
\end{align}
where
\begin{align}
\gamma & =  - \frac{\lambda}{4\Omega} \left[ \Omega-(\omega_0- \omega)\right]\\
\Omega & = \sqrt{\lambda^2 + (\omega- \omega_0)^2}\,.
\end{align}
The physics underlying this approximated Hamiltonian is that of a system 
with avoided level crossing and a gap $\simeq 2\gamma$ separating the 
otherwise crossing unperturbed levels. The quantity $\Delta$ introduced 
in the previous Sections is here given by $\Delta = \Omega - 2\omega$. 
Inserting this expression in Eq. (\ref{qfi0}) we arrive at the QFI
\begin{align}
H(\lambda) \stackrel{\omega\simeq\omega_0}{\simeq} \frac{1}{64 \omega_0^2} 
\frac{1}{(1-y+\frac{17}{64} y^2)^2} \qquad y = \frac{\lambda}{\omega_0}\,,
\end{align}
where, for the sake of simplicity, we have reported only the expression
close to resonance $\omega\simeq\omega_0$. The QFI is maximised for $\lambda=
\frac{32}{17}\, \omega_0$, indicating that for any value of $\lambda$ optimisation 
may be achieved by tuning the natural frequency of the well. As proved in the 
previous Sections, those ultimate limits to precision may be achieved by measuring
any observable of the form $\sigma_\theta=\sigma_1 \sin\theta + \sigma_3\cos\theta$,
where $\sigma_3$ is here the population of the unperturbed levels and $\sigma_1$
the corresponding polarisation. More general driven systems with level anticrossing
\cite{dri2} may be also addressed in the same way.
\subsection{Effective description of three-level systems}
Level anti-crossing may also occur in systems with more than two 
levels. In this case, the additional levels may influence the 
form of the eigenstates and, in turn, the behaviour of the QFI
when the value of the parameter $\lambda$ is perturbed.
Let us consider a three-level system with two close energy levels
and a third level being well separated in energy and weakly coupled 
to the first two levels. The Hamiltonian for such a system 
reads as follows
\begin{align}\label{hamlam3}
{\cal H}^{(3)} & = \left(\begin{matrix} \omega_1 (\lambda) & \gamma(\lambda) & g \\ 
\gamma^* (\lambda) & \omega_2 (\lambda) & g \\ 
g & g & \epsilon \end{matrix}\right)  \,,
\end{align}
where we assume a large gap between the third level 
and the others, i.e. $\epsilon \gg \omega_k $ and a weak 
coupling $g \ll 1$. In this regime, the system is amenable to 
an effective two-level description \cite{wim11}, with an effective 
Hamiltonian given by 
\begin{align}\label{hamlam3a}
{\cal H}^{(2)}_{\hbox{\tiny eff}} & = \left(\begin{matrix} \omega_1 (\lambda) + g^2/\epsilon & \gamma(\lambda) + g^2/\epsilon  \\ 
\gamma (\lambda) + g^2/\epsilon& \omega_2 (\lambda) + g^2/\epsilon 
\end{matrix}\right)  \,,
\end{align}
where we have also assumed $\gamma\in {\mathbb R}$. Using this effective
description we may now exploit the approach of the previous Sections
in order to assess the performances of this system as a scheme to estimate
the value of the $\lambda$. The QFI may be evaluated using Eq. (\ref{qfi0}).
Up to first order in the quantity $\kappa = g^2/\epsilon$ we have
\begin{align}
H_\kappa (\lambda) = & H_0 (\lambda)  \\   & - 2 \kappa \sqrt{H_0(\lambda)}  
\,\, \frac{2 \gamma \Delta \partial_\lambda\gamma + \partial_\lambda\Delta (\Delta^2-\gamma^2)}{(\gamma^2+\Delta^2)^2} \notag 
\end{align}
where $H_0 (\lambda)$ is the QFI of Eq. (\ref{qfi0}), corresponding 
to $\kappa=0$, i.e. a genuine two-level system. The possibility 
of enhancing estimation by coupling with additional levels is thus 
depending on the explicit dependence on $\lambda$ of the quantities 
$\gamma$ and $\Delta$.
\section{Conclusions}
\label{s:out}
Systems with Hamiltonian depending on a single parameter exhibits 
level anti-crossing. In turn, small perturbations to the value of 
the parameter may induce relevant changes in the system ground state,
which may be detected by measuring some accessible observable. 
Level anti-crossings may thus represent a resource for the 
characterization of Hamiltonians and for parameter estimation.
\par
Here, we have addressed in details metrological applications 
of level anti-crossing and have shown that universally optimal strategies 
for parameter estimation may be designed, independently on the nature and 
the value of the parameter of interest. In particular, we have studied 
quantum estimation for parameter dependent two-level Hamiltonians and show 
analytically that universal optimal estimation is achievable.
We also found that universally optimal estimation may be achievable 
also at high temperature if only the transverse, or only the diagonal, 
part of the Hamiltonian depends on the parameter. 
\par
We have also analyzed few examples, which confirm the generality of 
our approach and pave the way for further applications.
\begin{acknowledgments}
The authors thank C. Benedetti for stimulating discussions in the 
earlier stage of this work. 
\end{acknowledgments}


\begin{thebibliography}{99}
\bibitem{vnw} J. von Neumann, E.P. Wigner, Z. Physik, {\bf 30}, 467, (1929).
\bibitem{raz72} K. Razi Naqvi, Chem. Phys. Lett. {\bf 15}, 634, (1972).
\bibitem{lac1} T. H. Schukan, H. A. Weidenmiiller, Ann. Phys. {\bf 73}, 108 (1972).
\bibitem{wok81} A. Wokaun, S. C. Rand, R. G. DeVoe, R. G. Brewer, Phys. Rev. B {\bf 23}, 5733 (1981).
\bibitem{lac2} E. D. Davis, W. D. Heiss, J. Phys. G, {\bf 12}, 805 (1986).
\bibitem{hei90} W. D. Heiss, A. L. Sannino, J. Phys. A {\bf 23}, 1167 (1990).
\bibitem{prc1} H. C. Longuet-Higgins,Proc. R. Soc. Lond. A {\bf 344} 0095 (1975).
\bibitem{zphy} Q.-L. Jie, S.-J. Wang, L.-F. Wei, Z. Phys. B {\bf 104}, 373 (1997).
\bibitem{zph1} Q.-L. Jie, G. O. Xu, Comm. Th. Phys. {\bf 27}, 41 (1997).
\bibitem{res2} L. A. Collins, B. I. Schneider, C. J. Noble, C. W. McCurdy, 
S. Yabushita, Phys. Rev. Lett. {\bf 57}, 980 (1986).
\bibitem{res1} T. Timberlake, L. E. Reichl, Phys. Rev. A {\bf 64}, 
033404 (2001).
\bibitem{cha0} B. Eckhardt, Phys. Rep. {\bf 163}, 205 (1988).
\bibitem{cha3} M. Latka, P. Grigolini, B. J. West, Phys. Rev. E {\bf 50}, 596
(1994).
\bibitem{cha2} F. J. Arranz, F. Borondo, R. M. Benito, J. Chem. Phys. 
{\bf 107}, 2395 (1997). 
\bibitem{cha1} V. Ahlers, R. Zillmer, A. Pikovsky, Phys. Rev. A {\bf 63}, 
036213 (2001).
\bibitem{molma} F. Troiani, M. G. A. Paris, Phys. Rev. B {\bf 94}, 115422 (2016).
\bibitem{wic16} A. Wickenbrock, H. Zheng, L. Bougas, N. Leefer, S. Afach, A. Jarmola, 
V. M. Acosta, and D. Budker, Appl. Phys. Lett. {\bf 109}, 053505 (2016).
\bibitem{pars0} A. Fujiwara, H. Imai, J. Phys. A {\bf 36}, 8093 (2003).
\bibitem{pars1}
A. Sergeevich, S. D. Bartlett in {\em WCCI IEEE World Congress on Computational Intelligence} (2012)
\bibitem{pars5} B. Teklu, S. Olivares, M. G. A. Paris, J. Phys. 
B \textbf{42}, 035502 (2009).
\bibitem{pars4}
D. Brivio, S. Cialdi, S. Vezzoli, B. Teklu, M. G.
Genoni, S. Olivares, M. G. A. Paris, Phys. Rev. A \textbf{81}, 012315 (2010).
\bibitem{pars6} S.  Schirmer, G. Sophie, F. C. Langbein, Phys. Rev. A {\bf 91}, 
022125 (2015).
\bibitem{pars2} J. Suzuki, Int. J. Quantum Inf. {\bf 13}, 1450044 (2015).
\bibitem{pars3} J. Suzuki, J. Math. Phys. {\bf 57}, 042201 (2016).
\bibitem{hlb} C. W. Helstrom, {\em Quantum Detection and Estimation Theory} 
(Academic Press, New York, 1976).
\bibitem{lqe} M. G. A. Paris, Int. J. Quant. Inf. {\bf 7}, 125 (2009).
\bibitem{e1} A. Zwick, G. A. Alvarez, G. Kurizki, Phys. Rev.  Appl. {\bf 5}, 014007 (2016).
\bibitem{e2} A. Monras and M. G. A. Paris, Phys. Rev. Lett. {\bf 98}, 160401 (2007).
\bibitem{pinel13}O. Pinel, P. Jian, N. Treps, C. Fabre, and D. Braun, Phys. Rev. A {\bf 88}, 040102(R) (2013).
\bibitem{e3} M. G. Genoni, S. Olivares, M. G. A. Paris, Phys. Rev. Lett {\bf 106}, 153603 (2011).
\bibitem{e4} G. Brida, I. P. Degiovanni, A. Florio, M. Genovese, P. Giorda, A. Meda, M. G. A. Paris, and A. Shurupov, Phys. Rev. Lett.
{\bf 104}, 100501 (2010).
\bibitem{e5} G. Brida, I. P. Degiovanni, A. Florio, M. Genovese, P. Giorda, A. Meda, M. G. A. Paris, and A. P. Shurupov, Phys. Rev. A {\bf 83}, 052301 (2011).
\bibitem{e6} R. Blandino, M. G. Genoni, J. Etesse, M.
Barbieri, M. G. A. Paris, P. Grangier, and R. Tualle-Brouri, Phys. Rev. Lett. {\bf 109}, 180402 (2012).
\bibitem{e7} C. Benedetti, A.P Shurupov, M. G. A.
Paris, G. Brida, and M. Genovese, Phys. Rev. A {\bf 87}, 052136 (2013).
\bibitem{e8} A. Monras, Phys. Rev. A, {\bf 73}, 033821 (2006). 
\bibitem{e9} M. Kacprowicz, R. Demkowicz-Dobrzanski, W. Wasilewski, K. Banaszek, I. A. Walmsley, Nat. Phot. {\bf 4}, 357 (2010). 
\bibitem{e10} N. Spagnolo, C. Vitelli, V. G. Lucivero, V. Giovannetti, L. Maccone, and F. Sciarrino,  Phys. Rev. Lett. {\bf 108}, 233602 (2012).
\bibitem{e11} M. Brunelli, S. Olivares, and M. G. A. Paris, Phys. Rev.  A {\bf 84}, 032105 (2011).
\bibitem{e12} L. A. Correa, M. Mehboudi, G. Adesso, A. Sanpera, 	Phys. Rev. Lett. {\bf 114}, 220405 (2015).
\bibitem{e13} M. P. V.  Stenberg, Y. R. Sanders, and F. K. Wilhelm
Phys. Rev. Lett. {\bf 113}, 210404 (2014).
\bibitem{e14} M. Bina, I. Amelio, and M. G. A. Paris, 
Phys. Rev. E {\bf 93}, 052118 (2016).
\bibitem{e15} D. Tamascelli, C. Benedetti, S. Olivares, and M. G. A. Paris, 
Phys. Rev. A {\bf 94}, 042129 (2016).
\bibitem{e16} J. Nokkala, S. Maniscalco, J. Piilo, arXiv:1708.09625.
\bibitem{fid1} M. Cozzini, P. Giorda, P. Zanardi, Phys. Rev. B {\bf 75}, 
014439 (2007).
\bibitem{fid2} P. Zanardi, M. Cozzini, P. Giorda, J. Stat. Phys. {\bf 2007}, 
L02002 (2007).
\bibitem{fid3} P. Giorda, P. Zanardi, Phys. Rev. E {\bf 81}, 017203 (2010).
\bibitem{wim11} P. Plotz, M.  Lubasch, S. Wimberger, Physica A {\bf 390}, 1363
(2011).
\bibitem{geoNG} {M. G. Genoni, P. Giorda, M. G. A. Paris},
J. Phys. A \textbf{44}, 152001 (2011).
\bibitem{crb} H. Cram\`er, {\em Mathematical methods of statistics}, 
(Princeton University Press, 1946).
\bibitem{b1}
C. Benedetti, M. G. A. Paris, Phys. Lett. A {\bf 378}, 2495 (2014).
\bibitem{b2}
C. Benedetti, F. Buscemi, P. Bordone, M. G. A. Paris, Phys. Rev. A {\bf 89}, 032114 (2014).
\bibitem{b3}
M. G. A. Paris, Physica A {\bf 413}, 256 (2014).
\bibitem{b4}
M. A. C. Rossi, M. G. A. Paris, Phys. Rev. A {\bf 92}, 010302(R) (2015).
\bibitem{mug09} I. Lizuain, E. Hernandez-Concepci\`on, J. G. Muga, Phys. Rev. A {\bf 79}, 065602 (2009).
\bibitem{jchem}
M. A. C. Rossi, M. G. A. Paris, J. Chem. Phys. {\bf 144}, 024113 (2016).
\bibitem{holt93}
M. Holtaus, D. Hone, Phys. Rev. B {\bf 47}, 6499 (1993). 
\bibitem{dri2} B. P. Holder, L. E. Reichl, Phys. Rev. A {\bf 72}, 
043408 (2005).
\end{thebibliography}
\end{document}